\documentclass[runningheads]{llncs}
\usepackage{multirow}
\usepackage{graphicx}
\title{Statutory Professions in AI governance and their consequences for explainable AI\thanks{This paper emanated from research funded by Science Foundation Ireland to the Insight Centre for Data Analytics (12/RC/2289\_P2) and SFI Centre for Research Training in Machine Learning (18/CRT/6183). For the purpose of Open Access, the author has applied a CC BY public copyright licence to any Author Accepted Manuscript version arising from this submission.}}
\titlerunning{Statutory Professionals in AI Governance}
\author{Labhaoise NiFhaolain\inst{1}\orcidID{0009-0009-3788-531X} and Andrew Hines\inst{1}\orcidID{0000-0001-9636-2556}and Vivek Nallur\inst{1}\orcidID{0000-0003-0447-4150}}
\authorrunning{L. NiFhaolain et al.}
\institute{School of Computer Science, University College Dublin, Ireland \\
\email{labhaoise.ni.fhaolain@gmail.com},
\email{andrew.hines@ucd.ie},
\email{vivek.nallur@ucd.ie}}
\begin{document}
\maketitle              

\begin{abstract}
Intentional and accidental harms arising from the use of AI have impacted the health, safety and rights of individuals. While regulatory frameworks are being developed, there remains a lack of consensus on methods necessary to deliver safe AI. The potential for explainable AI (XAI) to contribute to the effectiveness of the regulation of AI is being increasingly examined.  Regulation must include methods to ensure compliance on an ongoing basis, though there is an absence of practical proposals on how to achieve this. For XAI to be successfully incorporated into a regulatory system, the individuals who are engaged in interpreting/explaining the model to stakeholders should be sufficiently qualified for the role. Statutory professionals are prevalent in domains in which harm can be done to the health, safety and rights of individuals. The most obvious examples are doctors, engineers and lawyers. Those professionals are required to exercise skill and judgement and to defend their decision making process in the event of harm occurring. We propose that a statutory profession framework be introduced as a necessary part of the AI regulatory framework for compliance and monitoring purposes. We will refer to this new statutory professional as an AI Architect (AIA). This AIA would be responsible to ensure the risk of harm is minimised and accountable in the event that harms occur. The AIA would also be relied on to provide appropriate interpretations/explanations of XAI models to stakeholders. Further, in order to satisfy themselves that the models have been developed in a satisfactory manner, the AIA would require models to have appropriate transparency. Therefore it is likely that the introduction of an AIA system would lead to an increase in the use of XAI to enable AIA to discharge their professional obligations.  

\keywords{Artificial Intelligence \and XAI \and Governance  \and Regulation \and Statutory Profession.}
\end{abstract}
\section{AI Regulation and Challenges}
Intentional and accidental harms arising from the use of AI have impacted the health, safety and rights of individuals~\cite{Acemoglu2021Harm}. While there is much pre-existing legislative regulation applicable to AI~\cite{Viljanen2022AI}, and AI specific regulatory frameworks are being developed~\cite{EUCommission2021Proposal}, there subsists a lack of consensus on the wider governance framework necessary to deliver safe AI~\cite{Smuha2021From}. An effective governance framework must include mechanisms to ensure compliance on an ongoing basis. To date, in terms of compliance mechanisms in AI regulation, the focus has been on seeking to make corporations and companies responsible and accountable. However, these entities are explicitly structured through incorporation to avoid personal liability~\cite{Easterbrook1985Limited}. Gaps in accountability in AI have been reviewed~\cite{Santoni2021Four,Hohma2023Investigating} and some work has been carried out on the dearth of practical proposals on how to move from principles to practice in AI governance~\cite{Palladino2021Biased}. Human oversight in relation to AI has been explored~\cite{Methnani2021Let,Santoni2018Meaningful} as has the empowerment of employees through continuous education and change management to operationalise AI governance~\cite{Mokander2022Challenges}. However there remains a lack of proposals on how to structure a system of accountability and associated enforcement that interacts with the legal system in a pragmatic manner. In effect, the operationalisation of accountability and attendant enforcement remains absent.   The potential for XAI to contribute to the effectiveness of the governance of AI within the legal system is being increasingly examined~\cite{Ebers2021Regulating}. In this paper we seek to establish the potential for combining XAI with personal obligations that underpin statutory professions to illustrate how they can contribute to the enforceable accountability for AI through regulation.

Regulation seeks to modify behaviour to achieve particular outcomes~\cite{Shaffer2010Hard} and a regulatory framework encompasses norms along with monitoring and correction.  While an AI regulatory framework is far from complete,  regulatory norms for AI exist including ethics principles and codes of conduct~\cite{Council2020AIInitiatives}, standards~\cite{Winfield2019Ethical} and law (both applying existing law and proposals for Regulations~\cite{EUCommission2021Proposal}).  These are all necessary components but are not sufficient. Operational and adherence mechanisms are also required.  Voluntary guidelines, such as ethics principles and codes of conduct, require deep embedding for success~\cite{Webley2008Corporate}, and rely on human actors. Standards require government oversight to deliver societal benefit~\cite{Garvin1983Can}. For compliance with standards, Huising and Silbey~\cite{Huising2011Governing} argue that individuals’ conduct and personal responsibility is crucial for implementation. By design, corporate accountability limits personal responsibility and, in order to minimise harm, a regulatory system needs individuals who are responsible for preventing harm, and held accountable if harms do occur.  

Regulation of professionals  arises predominantly  in domains with significant impact on health, safety and rights due to the fundamental importance of these areas to the individual and to society as a whole. While lawyers and doctors have been the subject of some form of governance for millenia, modern-day professions emerged in the late nineteenth and early twentieth centuries~\cite{Law2005Specialization}.  Statutory professions are those whose conduct is regulated by legislation, who can be subject to the rules (including educational requirements) and codes of conduct set out by a regulatory body, and who can be disciplined by the regulatory body and the judiciary for infractions.  The EU Commission has defined a regulated profession as “A professional activity or group of professional activities, access to which, the pursuit of which, or one of the modes of pursuit of which is subject, directly or indirectly, by virtue of legislative, regulatory or administrative provisions to the possession of specific professional qualifications.” ~\cite{EUParliament2005Recognition}. In some jurisdictions the use of certain titles (known as \textit{reserved titles}) is limited. Prerequisites for using the title may include prescribed qualifications and registration with the relevant body. However this falls short of a fully statutory profession. For example, the title of \textit{software engineer} is a reserved one in every province and territory in Canada. Each of the ten provinces and three territories regulates the work of software engineering in a different way, under their own legislative provisions. Therefore while the title is limited in a uniform manner, the work is not controlled in a similar way.  Indeed, it has been noted by Engineers Canada, an national umbrella body for provincial and territorial regulators, that the regulation of software engineering practice has been troublesome due to its overlap with software development~\cite{Engineers2023Professional}. 

In considering whether to create a new statutory profession, the governing question asked in most jurisdictions is whether there is the risk of harm to the public as a result of the professional's conduct~\cite{Mills2023Disciplinary}. In other domains, where individuals can be negatively affected by the conduct of service providers, those service providers are regulated.  However, whilst computer scientists and software developers can impact the health, safety and rights of individuals across society when developing AI systems, individuals are not regulated. 

We propose that a statutory profession framework, akin to the system for lawyers, doctors and engineers, be introduced as a necessary part of the AI regulatory framework, for compliance and monitoring purposes.  We will refer to this statutory professional as an AI Architect (AIA). In this paper the term AIA is simply used as a working title for the proposed professional and the specific term is not of significance at this juncture.  

\section{XAI Interface Challenges}
There is much discussion about \textit{humans in the loop, on the loop} and \textit{in command} and about the importance of the role from a legal perspective, though there is no legal definition of what these terms entail~\cite{Enarsson2022Approaching}. 
Indeed the \textit{human in the loop} is “no panacea” for the concerns arising from AI~\cite{Krugel2023Algorithms}, but as governance is a social structure, humans are central it.

The goal of XAI is to produce human-interpretable models~\cite{Ali2023Explainable} and to \textit{justify} decisions which have been made, particularly when used in critical situations~\cite{Meske2022Explainable}.

Ali et al.~\cite{Ali2023Explainable} categorise XAI techniques as follows: data explainability, model explainability, post-hoc explainability and assessment of explanations.  The authors also outline the role of humans in the assessment of the explanations and set out post-hoc explainability methods as follows: attribution methods, visualisation methods, example based explanation methods, game theory methods, knowledge extraction methods and neural methods. These techniques are used by people in order to arrive at an explanation - therefore these methods can only be deployed in the context of a \textit{human-computer interaction}~\cite{Ali2023Explainable}. 

Miller et al. \cite{Miller2017Explainable} have identified that moving from technical aspects to the mindset of the user can be challenging for XAI designers. In proposing a process for the design of user-centric XAI systems as a means to avoid unwanted consequences, Förster et al.~\cite{Forster2020Fostering} recommend a higher user involvement in the process. 

Given that the veracity of explanation is at the core of the future of XAI, challenges arise when explanations are provided which are guided by the explainer’s agenda, rather than the individual impacted by the decision.  Reference has been made  to the  rise of  \textit{altered} explanations in the field of XAI which pose problems in the field~\cite{Chromik2019Dark,Schneider2022Deceptive}. Examples of \textit{altered} explanations being provided include in order to avoid intellectual property disclosure, or a bank employee denying loans for nefarious reasons, or companies using explanations to lure customers~\cite{Schneider2022Deceptive}.  

Doran et al.~\cite{Doran2018What} have drawn the distinction between how “rules” of certain XAI systems can shed light on \textit{how} decisions are made but they do not explain \textit{why} the decision was made. The authors  state that the annotations and visualisations produced within the XAI domain require “human-driven post processing under their own line of reasoning”. 

\begin{table}[t]
\caption{Summary of regulated professions in Ireland with a requirement for professional indemnity insurance categorised by risk} 
\resizebox{\columnwidth}{!}{%
\begin{tabular}{lll}
\textbf{Health}             & \textbf{Rights}                                 & \textbf{Safety}     \\
\hline\\
\multirow{16}{*}{\parbox{9cm}{Medical (Anaesthesiology, Basic medical training - Ireland, Cardiology, Cardiothoracic surgery, Chemical pathology, Child and adolescent psychiatry, Clinical genetics, Dermatology, Emergency medicine, Endocrinology and diabetes mellitus, Gastro-enterology, General (internal) medicine, General Medical Practitioner, General surgery, Genito-urinary medicine, Geriatric medicine, Haematology (clinical and laboratory), Histopathology, Immunology (clinical and laboratory), Infectious diseases, Medical oncology, Microbiology, Nephrology, Neurology, Neurosurgery, Obstetrics and gynaecology, Occupational medicine, Ophthalmic surgery, Ophthalmology, Oral and maxillo-facial surgery, Otolaryngology, Paediatric Surgery, Paediatrics, Pharmaceutical Medicine, Plastic, reconstructive and aesthetic surgery, Psychiatry, Public health medicine, Radiation oncology, Radiology, Respiratory medicine, Rheumatology, Trauma and orthopaedic surgery)}} &
  Auctioneer &
  Aeromedical Examiner (AME) \\
                   & Building Energy Rating (BER) Assessors & Affiliate (engineering) Associate Engineer         \\
                   & Certified Public Accountant            & Architect                                          \\
                   & Chartered Accountant/Auditor           & Architect (acquired rights)                        \\
                   & Chartered Certified Accountant         & Building Surveyor                                  \\
                   & Chartered Tax Adviser                  & Chartered engineer                                 \\
                   & Estate Agents                          & Engineering Technician                             \\
                   & Incorporated Public Accountant         & Quantity Surveyor                                  \\
                   & Letting Agents                         & Registered Electrical Contractor                   \\
                   & Liquidator                             & Transport Manager for a Road Transport Undertaking \\
                   & Management Agent                       &                                                    \\
                   & Personal Insolvency Practitioner       &                                                    \\
                   & Solicitor                              &                                                    \\
                   &                                        &                                                    \\
                   &                                        &                                                    \\
                   &                                        &                                                    \\
Veterinary Surgeon &                                        &                                                   
\end{tabular}%
}
\end{table}

\section{Regulated Professions in the EU }
At a general level, there are different elements in systems of regulated professions. The professional has a legal/regulatory requirement to register with a prescribed body (regulatory body). The regulatory body is charged with assessing whether the individual has reached a requisite level of education/qualification (at admission to the profession and often on an ongoing basis) and the professional is subject to rules which govern their conduct (in their professional role, and sometimes outside of that role). If the professional breaches the rules of conduct, the regulatory body (usually in conjunction with the Court) can prevent the professional from practising in that regulated role. 

Professional indemnity (PI) Insurance can provide indemnity against claims made by those impacted by the professional’s conduct.  It is designed to compensate for losses which have arisen from a professional’s failures and is not entirely dependent on the professional’s own financial assets~\cite{Morgan2017Professional}.
Morgan and Hanrahan~\cite{Morgan2017Professional} have set out the overall purpose of PI insurance, stating that “The reasons for mandating PI insurance have public interest elements, encompassing both the protection of individual clients in the event of professional failure that results in a loss, and the creation and support of trust in the general body 
of professionals”. Cannon and McGurk~\cite{Cannon2016Professional} have noted that the purpose of mandating this type of insurance is to protect the clients and those impacted by the professional’s conduct, rather than protecting the professional. 

Therefore society requires certain professionals to carry PI insurance due to their potential to cause serious damage to individuals. In order to examine the relevance  for a regulated profession to enhance the governance of AI, we conducted a survey of regulated professions in Ireland as set out by the European Commission in the Regulated Professions Database~\cite{EUCommission2023Database}. That database contains the regulated professions of all EU countries and provides a search and filter function. Using web data extraction we filtered the results to include regulated professions from Ireland only. We then identified those professions for whom PI insurance is a prerequisite for practising in the area (e.g. solicitor~\cite{EUCommission2023Solicitor}).

 The total number of regulated professions in Ireland which appear on the European Database of Regulated Professions~\cite{EUCommission2023Database} is 229. Of those, 69 are required to carry PI insurance.  We have set out these professions in Table 1. We examined that subsection and categorised them. The three risk areas into which these professions fall are very clearly identifiable as those of health, safety and rights. The impact on rights includes an impact on property rights (whether real property in the form of land and buildings or other property). It also includes the civil and political rights, along with economic, social and cultural rights dealt with by solicitors.

\begin{figure}[ht]
    \centering
    \includegraphics[width=\textwidth]{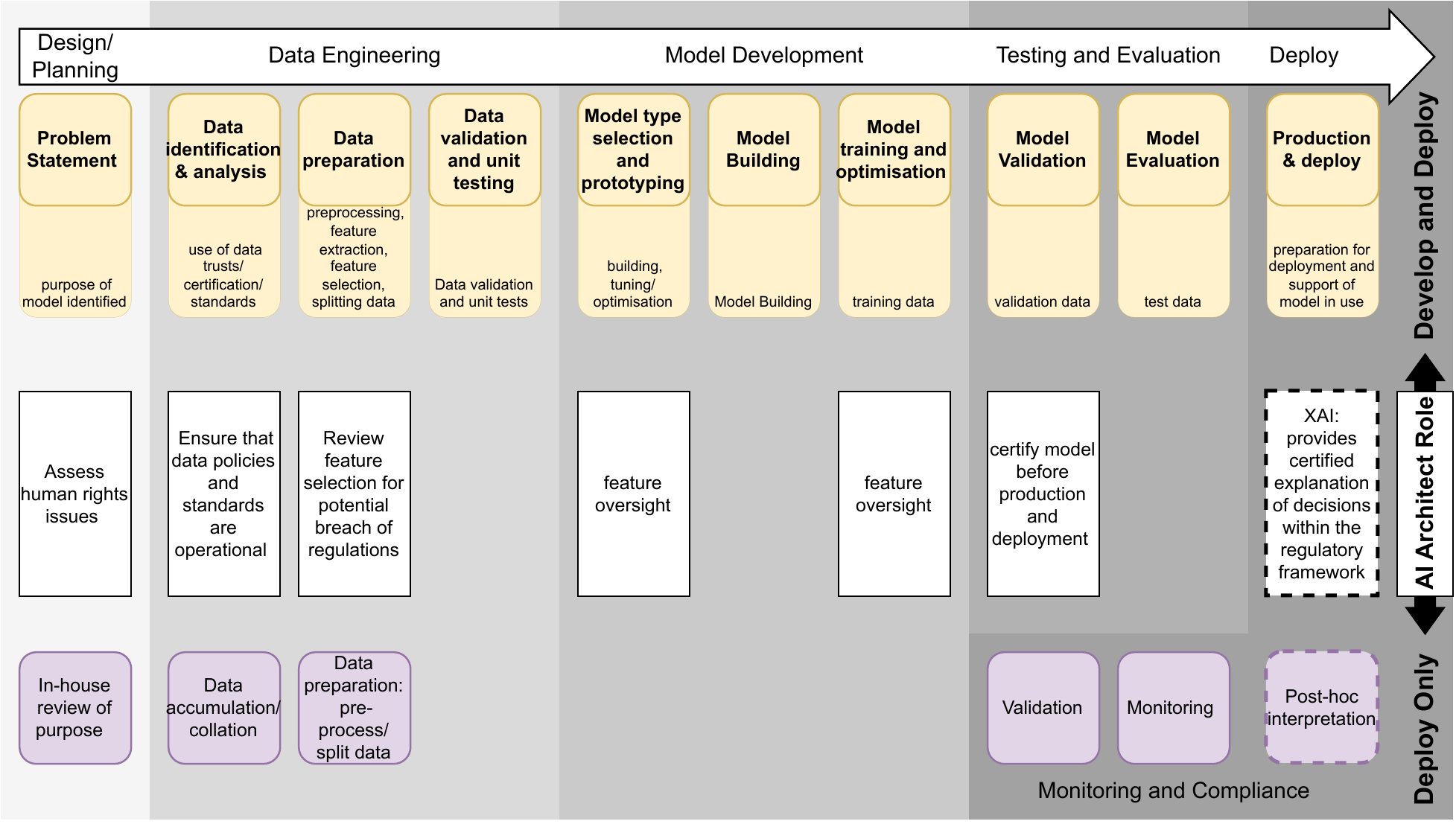}
    \caption{Suggested functions of AI Architect in development of ML solution and in deployment of ML solution.}
    \label{fig:my_label}
\end{figure}

\section{Proposal for AI Architect to Address Regulatory and XAI Challenges}
We propose that AI regulation should be operationalised effected through a human actor, specifically by requiring through legislation, that certain AI solutions be ‘signed off’ by an AIA who is a member of a statutory profession, as previously proposed~\cite{NiFhaolain2020Could}. In the case of XAI, the AIA would be the person entitled to provide and sign off on the interpretation/explanation of the model. Like other domains, not all those working in computer science or software engineering would require regulation within a statutory profession, but AIA would be a protected title for those with the responsibility and accountability for sign-off. We have set out in Figure 1. an outline of a Machine Learning Operations (ML Ops) pipeline. The top row sets out the instance of development and deployment of an ML solution, while the bottom row sets out deployment only. The middle row illustrates how the pipeline is punctuated with the involvement of an AIA. The AIA's role would be explicitly outlined and tightly coupled to the AI system's lifecycle. In this way, the AIA's obligations would be clearly delineated and bounded.  The dotted blocks are references to the role of the AIA in the case of XAI, in respect of providing explanations of XAI within the regulatory framework. As can be seen from Figure 1., the AIA role may be wide ranging. Much like the profession of~\textit{doctor} has evolved into sub-specialties, as can be seen from the first column in Table 1., a similar evolution would likely occur in the case of the AIA profession. Indeed, the field of AI engineering as a discipline is already being explored and is gaining momentum~\cite{Carnegie2023Artificial,Ozkaya2022AI}.

The AIA would have a professional obligation to ensure that AI solutions being developed and deployed are safe, by ensuring that due diligence is performed on datasets, following best practices, and remaining up to date with technological developments. Crucially, professional obligations (given their statutory status) would take precedence over employee obligations. Tensions may arise between the desires of the company and the requirements of the regulations governing the AIA's conduct. Similar tensions arise in the case of solicitors who are employed by companies as \textit{in-house solicitors}, where the companies are in effect their sole clients. Like solicitors, the AIA would be required to conduct themselves according to stipulated principles. The Solicitors Regulation Authority of England and Wales sets out the principles required of solicitors which include independence, honesty and integrity. In the event that the principles come into conflict with the client's interests, the principles which safeguard the wider public interest take precedence~\cite{SRA2018Principles}.  
The question arises of the employment security of an AIA if they are obliged to make \textit{unpopular} decisions. The AIA would benefit from the protection of the regulatory body system such that they cannot be dismissed from their employment for carrying out these professional and legally protected obligations. Any attempt to dismiss an employee could be deemed to be wrongful dismissal. This combined obligation and protection dynamic balances the power between individual professionals and large corporations.  

Personal responsibility would be a component of the system for AIAs. This would not be a means to absolve corporations of responsibility, rather a recognition that the very rationale of the modern corporation is to avoid personal liability. This would address the ongoing challenge of making shareholders liable for the obligations of their corporation as \textit{piercing the corporate veil} of limited personal liability is fraught with obstacles \cite{Muchlinski2010Limited}. 

In order to deliver personal responsibility and accountability, legislation and codes of conduct exist by which the professionals must abide. Failure to comply can result in disciplinary procedures which are instigated by the errant professional's regulatory body. Mills et al.~\cite{Mills2023Disciplinary} have stated that “Professional-specific grounds for misconduct reflect the approach that professionals are, if the profession itself is to be trusted and well regarded, to be held to a different – typically higher – standard than non-professionals". In broad outline, following successful disciplinary proceedings, sanctions may be brought against the professional. The ultimate sanction, usually reserved for a disciplinary procedure which comes before a judge in court, is to strike the individual from the register of professionals, preventing them from practising in that professional role. Disciplinary sanctions can also have cross border implications. Under the EU Direction on the recognition of professional qualifications, certain professional regulators must notify parallel competent authorities in other member states when certain individual professionals have been restricted or prohibited in their practice~\cite{EUParliament2005Recognition}.

In parallel with the disciplinary proceedings, legal claims for professional negligence can arise, this time brought by those who have been harmed by the conduct of the professional. In these cases the professional may make seek indemnity from their professional indemnity insurers. This twin track approach (a) protects society in general through the disciplinary process (through sanctions deterring or preventing the professional from causing similar harm again) and (b) restores the individual to the position they were in before the harm caused by the professional occurred (through an award of damages or through another restorative court order). Whether in the context of disciplinary proceedings or professional negligence claims, the professional is obliged to justify their conduct and the decisions they made to the satisfaction of experts in that field, rather than lay people. 

In the case of AI, the role of AIA would be different from that of \textit{ethical lead} and similar roles within AI development and deployment. Currently, without the statutory footing, it is open to companies to change their internal ethics guidelines or choose to disregard external ethics guidelines or standards, due to their voluntary nature. This can result the role of “ethical leads" being rendered ineffective, or indeed can result in the role being made redundant through termination of employment contracts. 

A search of the internet will produce many results relating to certification of computer scientists including some calls for \textit{regulation} of that role. To the best of our knowledge there are no specific, measurable and realistic proposals for exactly how this should be done. Our proposal calls for a fully regulated profession with statutory obligations for professional conduct, liability, indemnity insurance and disciplinary regimes which can result in the professional being before the Court for breaches of their duties and being prevented from practising in the role. 

The AIA, as the regulated a human in the loop, at each stage of the development and deployment process would increase consumer/citizen confidence. This \textit{farm to fork} style responsibility tracing would require that an AIA oversees substantive changes in the solution, or the deployment environment. An AIA system would benefit those corporations that see value in a market with legal certainty, while nefarious actors would be disadvantaged. Given that fast moving fields such as AI are difficult to legislate for, an agile AIA system would also facilitate better State protection of citizens. 

The individual fulfilling the AIA role would have personal responsibility to ensure that explanations provided in XAI are accurate, rather than being driven by the deploying company. This would avoid the problem of \textit{altered} explanations, referred to above. In addition, in the context of an increase in the interactions between technical teams and stakeholders in the development phases of XAI, those interactions could be mediated by the AIA, who can adopt a neutral stance between the two cohorts. The characteristics of XAI would mean that AIAs could more confidently ensure that the AI systems are developed in compliance with the relevant regulations.  Therefore it is likely that the introduction of this regulated profession would garner an increase in the use of XAI techniques.

The AI Act may be a natural route through which to introduce the AIA into the regulatory system.  Recital 48 of the draft AI Act compromise text requires “human oversight" in particular scenarios and requires that “natural persons to whom human oversight has been assigned have the necessary competence, training and authority to carry out that role". Article 14 requires high risk systems to have effective human oversight to prevent or minimise risks to health safety and rights of of individuals.  Under Article 29 the deployers of AI systems must “ensure that the natural persons assigned to ensure human oversight of the high-risk AI systems are competent, properly qualified and trained, and have the necessary resources in order to ensure the effective supervision of the AI system in accordance with Article 14". The requirements for human oversight would be fulfilled by an AIA while the transparency obligations under Article 13 may see an expansion of the use of XAI.

Many standards organisations continue their work in the field of AI~\cite{Turing2023Observatory}. While this form of governance has been criticised for being driven by industry, this has been acknowledged by the EU, resulting in a new strategy on Standardisation~\cite{EUCommission2022EU}. The AI Act makes explicit reference to the  use of standards and the EU Commission has issued a draft standardisation request to the European Standardisation Organisations for standards in support of safe and trustworthy AI~\cite{EUCommission2022Draft}. This draft request mandates CEN-CENELEC to include human rights expertise and civil society involvement in the development of the standards. As with any standards, a method to ensure that they are being complied with will be necessary. Again, the AIA would contribute to ongoing monitoring and compliance rather than evaluating compliance ex-post. 

The approach to bring the regulated profession of AIA into being would be multipronged.  The process would include the creation of a regulatory body and a protected title through legislation~\cite{Huber2013Should}. That body would work to set out the educational requirements and the practices which would be reserved to the regulated professional. The introduction of an AIA may be unpalatable to technology companies as a costly overhead. However much like regulation in Pharma is a cost of doing business, so too would the inclusion of an AIA.  The introduction of the professional to the market could be achieved through the requirement that an AIA be used in all cases of government procurement initially. This would create an incentive for corporations to bring AIAs into their operations. Phased introduction into other spheres could then proceed. 

Regulatory impact assessments (RIAs) are used as evaluation tools for policy, laws and regulations at national~\cite{Department2009Revised}, cross-national~\cite{Commission2021aBetter} and international levels~\cite{OECD2020Regulatory}. These RIAs would be built into the regulatory framework to allow transparent assessment during the evolution of the profession within the wider regulatory context.  Examples of ML pipelines which resulted, for example, in human rights breaches could be contrasted with a pipeline in which an AIA had defined tasks to be performed to consider whether the inclusion of the AIA reduced the incidence human rights breaches. 

The statutory professions system is not a perfect solution but its longevity in the areas of medicine, law and civil engineering demonstrates its ability to deliver on the objectives of service and product quality standards and to adapt over many centuries to changing circumstances. 

\section{Conclusion}
Governments around the world are signalling an appetite for AI regulation in AI strategies~\cite{NiFhaolain2020Assessing} and the EU has initiated work on a regulatory system~\cite{EUCommission2021Proposal}. When quality levels and rules are being determined, through ethics guidelines, standards and legislation, it is critical that monitoring and compliance mechanisms are developed in parallel. We propose that the inclusion of a statutory profession is a necessary, yet currently unconsidered, component of a sustainable and workable AI regulatory system. The concept of a computer science profession has been examined for decades~\cite{Holmes1974Social,Holmes2011Profession}. While some advances have been made, for example through being chartered as an IT professional, the profession has not evolved into a regulated profession with authority and responsibilities underpinned by legislative status. There are sufficient examples of effective statutory regulation of professionals in other similar domains to indicate that an AIA is a promising candidate for statutory regulation in order to enhance the regulation of AI.

\clearpage
\bibliographystyle{splncs04}

\end{document}